\documentclass[runningheads]{llncs}
\usepackage[T1]{fontenc}
\usepackage{graphicx}
\usepackage{booktabs}
\usepackage{multirow}
\usepackage{xcolor}
\usepackage{tabularx}
\usepackage{array}
\usepackage{url}
\usepackage{tikz}
\usepackage{pgfplots}
\pgfplotsset{compat=1.18}
\begin{document}
\title{NFR-to-Code Traceability in a Blockchain-IoT System: An Empirical Study}
\titlerunning{NFR-to-Code Traceability in a Blockchain-IoT System}
\author{Yifei Wang\inst{1} \and
Jacky Keung\inst{1} \and
Xiaoxue Ma\inst{2} \and
Shijie Zhang\inst{3} \and
Yishu Li\inst{2}\thanks{Corresponding author.}}
\authorrunning{Y. Wang et al.}
\institute{City University of Hong Kong, Hong Kong, China\\
\email{ywang4748-c@my.cityu.edu.hk, Jacky.Keung@cityu.edu.hk}
\and
Hong Kong Metropolitan University, Hong Kong, China\\
\email{\{kxma,sliy\}@hkmu.edu.hk}
\and
The Chinese University of Hong Kong, Shenzhen, China\\
\email{224040085@link.cuhk.edu.cn}}
\maketitle

\begin{abstract}
Requirement-to-Code traceability has been widely studied, yet existing research and public benchmarks remain largely centered on functional requirements (FRs). In contrast, traceability for non-functional requirements (NFRs) remains more difficult and underexplored, which hinders the verification of critical quality concerns such as security and reliability.
This paper studies NFR-to-Code traceability based on a real-world blockchain-IoT project. We design an annotation protocol for constructing trace links across heterogeneous artifacts and build a manually curated subset containing both FR and NFR links. Using this subset, we examine the characteristics of NFR traceability and further evaluate four representative retrieval baselines: TF-IDF, BM25, LSI, and WMD.
The results show that FR-to-Code tracing is consistently easier than NFR-to-Code tracing, while security-related NFRs are the most difficult subset. They further indicate that the main challenge of NFR traceability lies not in requirement availability, but in implementation evidence that is distributed and not clearly localized in code.

\end{abstract}
\section{Introduction}
Requirement-to-Code traceability refers to establishing explicit links between requirements and their implementing code elements~\cite{cleland2014software}. 
Such trace links support important software engineering tasks, including requirement validation, change impact analysis, regression testing, and compliance checking~\cite{rempel2016preventing}. More importantly, they play a critical role in software quality assurance by enabling the verification of whether key requirements, especially those related to security and reliability, are properly implemented in code.
However, most existing studies and datasets mainly focus on functional requirements (FRs)~\cite{abad2017works}, which are typically associated with explicit system behaviors and can be more directly mapped to identifiable implementation logic in code. In real-world systems, however, non-functional requirements (NFRs) are equally critical, especially for quality attributes such as security, performance, reliability, auditability, and data integrity~\cite{chung2012non}. 

NFRs are significantly harder to trace to code than FRs~\cite{cleland2005toward}. An FR typically maps to a relatively clear implementation entry point, such as a controller, service, or API logic. In contrast, NFRs are often implemented in a more distributed and implicit manner, with their evidence scattered across configuration settings, communication mechanisms, validation logic, performance constraints, testing artifacts, or maintenance-related components~\cite{cleland2005toward}. 
This difference in implementation characteristics is also reflected in existing traceability research and publicly available datasets, which have largely focused on FR-to-Code links, while support for NFR-to-Code traceability remains limited~\cite{mahmoud2016detecting,limaylla2023towards}.

To address this gap, this paper presents an empirical study of NFR-to-Code traceability based on a blockchain-IoT system. We first design an annotation protocol to construct trace links between requirement units and code entities across heterogeneous artifacts. Following this protocol, we build a manually curated trace subset containing both FR and NFR links over requirement documents, Java classes, and Solidity contracts.

Based on this dataset, we analyze the characteristics of NFR traceability across multiple categories and compare it with FR traceability, with a focused analysis of security-related NFRs.
To further support our analysis, we evaluate representative traceability baselines on the constructed dataset. 
The results indicate that NFR traceability is more challenging than FR traceability, especially for security-related NFRs due to their dispersed and implicit implementation patterns.

The main contributions of this paper include:
%% Contribution
\begin{itemize}
    \item We design an annotation protocol for constructing NFR-to-Code trace links in a real software project with heterogeneous requirements and code artifacts.
    \item We build a manually curated FR/NFR-to-Code trace subset for a blockchain-IoT system, covering requirement documents, Java classes, and Solidity contracts.
    \item We provide a qualitative analysis of NFR-to-Code traceability across categories, highlighting its challenges and differences from FR traceability, with a focus on security-related NFRs.
    \item We conduct an empirical comparison of FR-to-Code and NFR-to-Code traceability using representative retrieval baselines to support the observed challenges.
\end{itemize}

The rest of the paper is organized as follows. Section 2 introduces the study context; Section 3 presents the annotation protocol and the dataset; Section 4 reports the main observations and challenges from the dataset; Section 5 provides a baseline evaluation; Sections 6 and 7 discuss related work and threats to validity; and Section 8 concludes the paper.

\section{Study Context}
\subsection{Project Context}
This study is conducted on a real-world software project in a blockchain-IoT setting. The system supports IoT data acquisition, secure transmission, on-chain evidence storage, and trusted data verification. Its requirement-side artifacts contain both FRs (e.g., interface invocation and evidence storage) and NFRs related to security, performance, auditability, reliability, and data integrity.

A key characteristic of this project is the heterogeneity between requirement expression and implementation artifacts. On the requirement side, the available information is distributed across multiple project documents, primarily including a business requirement document and a technical specification, rather than a centralized requirement specification. On the implementation side, the codebase consists of a Java application with configuration, controller, service, and utility components, together with Solidity contracts for on-chain behavior. 

\subsection{Research Questions}
Based on the above context, this paper investigates the following research questions:
\begin{itemize}
\item \textbf{RQ1 } What is the distribution of NFRs in the curated subset in terms of categories and traceability outcomes?
\item \textbf{RQ2 } How does NFR-to-Code traceability differ from FR-to-Code traceability in terms of link distribution, implementation heterogeneity, and traceability difficulty?
\item \textbf{RQ3 } Which NFR categories appear more difficult to trace, and what implementation characteristics contribute to this difficulty?
\end{itemize}

\section{Dataset Construction and Annotation}
\subsection{Annotation Protocol}
The dataset is constructed through a structured annotation protocol designed to produce a manually curated reference subset for FR/NFR-to-Code traceability~\cite{ahmadiyah2023monthes,borg2014recovering}.
The protocol covers three main steps: (1) extracting and classifying requirement units from heterogeneous requirement documents, (2) defining file-level code units from the implementation artifacts, and (3) annotating trace links between requirement units and code units under explicit labeling criteria, with three annotation labels: \textit{Positive}, \textit{Negative}, and \textit{Missing}. 

To improve reliability, the annotation process involves two annotators, both senior developers from industry, who independently label the candidate requirement\allowbreak--code pairs by following the same protocol. Disagreements are then reviewed and resolved through discussion and adjudication. The final dataset is constructed from the agreed annotations after this conflict resolution process and is used in the subsequent descriptive analysis and baseline evaluation.

\subsection{Source Artifacts}
The source artifacts used in this study include both requirement-side and implementation\allowbreak-side artifacts, as shown in Table~\ref{tab:artifacts}. On the requirement side, the raw sources are primarily a business requirement document and a technical specification from the target project. These documents contain both functional descriptions and quality-related constraints, which are later normalized into traceable requirement units. Together, they form the raw natural-language basis for requirement unit extraction.
\begin{table}[t]
\centering
\caption{Raw and Derived Artifacts in the Curated Dataset}
\label{tab:artifacts}
\footnotesize
\setlength{\tabcolsep}{5pt}
\renewcommand{\arraystretch}{1.08}
\begin{tabular}{lll}
\toprule
\textbf{Artifact} & \textbf{Format} & \textbf{Description} \\
\midrule
Requirement documents & \texttt{doc/*.docx, *.doc} & Raw requirement sources \\
Requirement units & \texttt{requirements/*.txt} & Extracted units \\
Java code artifacts & \texttt{*.java} & Off-chain code units \\
Solidity code artifacts & \texttt{*.sol} & On-chain code units \\
Metadata & \texttt{requirements\_metadata.csv} & Requirement attributes \\
Trace links & \texttt{trace\_links.csv} & Annotated links \\
\bottomrule
\end{tabular}
\end{table}

On the implementation side, the artifacts consist of 22 file-level code units, including Java classes and Solidity contract files. The Java codebase contains configuration, controller, service, and utility components, while the Solidity files implement contract-level on-chain behavior. In this study, these file-level artifacts serve as the basic implementation units for trace construction.

In addition to the original documents and code artifacts, the study also produces structured derived artifacts, including extracted requirement units, requirement metadata, and adjudicated trace labels. These derived artifacts support the subsequent descriptive analysis and benchmark-style retrieval evaluation.

\subsection{Requirement Extraction and Classification}
Requirement units are derived from heterogeneous requirement documents using a structured extraction process.
Instead of treating full paragraphs or sentences as traceable elements, requirement descriptions are normalized into \textit{minimal verifiable requirement units}, each capturing an independently verifiable constraint or functional point. When a sentence contains multiple concerns, it is decomposed into multiple requirement units.

Each requirement unit is then classified as either an FR or an NFR. For NFRs, a quality category is further assigned according to the primary intent of the requirement, including \textit{Security, Performance, Reliability, Data Integrity, Auditability, Scalability,} and \textit{Maintainability}~\cite{chung2012non}. An example of requirement decomposition and classification is shown below.
\begin{center}
\fbox{%
\begin{minipage}{0.95\columnwidth}
\vspace{0.6em}
\small

\textbf{Requirement statement:}

\texttt{"The system shall provide secure device communication, ensure data integrity during transmission, and support timely response for common queries."}

\vspace{0.6em}

\textbf{Decomposed requirement units:}

R1: The system shall provide secure device communication.\\
\hspace*{1.5em}Category: NFR\\
\hspace*{1.5em}NFR\_Type: Security

\vspace{0.4em}

R2: The system shall ensure data integrity during transmission.\\
\hspace*{1.5em}Category: NFR\\
\hspace*{1.5em}NFR\_Type: Data Integrity

\vspace{0.4em}

R3: The system shall support timely response for common queries.\\
\hspace*{1.5em}Category: NFR\\
\hspace*{1.5em}NFR\_Type: Performance

\vspace{0.6em}
\end{minipage}%
}
\end{center}

Only explicit and verifiable system-level requirements are retained in the dataset. Background descriptions, business motivations, policy statements, and purely aspirational texts are excluded.

\subsection{Code Unit Definition}
Code units are defined at the \textit{file-level implementation entity} granularity. For the Java part of the system, each class is treated as a code unit, while for the blockchain component, each Solidity contract file is treated as a code unit. 
This definition follows the organization of the codebase and serves as the basis for trace construction.

This granularity balances traceability precision and annotation consistency. A finer-grained definition (e.g., methods or statements) would introduce ambiguity and annotation overhead, especially when implementation evidence is distributed across multiple methods within a class. A coarser-grained definition (e.g., packages or modules) would make trace links overly broad and reduce interpretability.

In addition to code unit identities, each artifact may be assigned a descriptive artifact type, such as \textit{Config, Controller, Service, Utility,} or \textit{Contract}, to facilitate later analysis. These labels are used only as supplementary metadata and do not alter the basic unit of trace construction.

\subsection{Trace Link Annotation}
Trace link annotation is guided by the principle of \textit{direct implementation responsibility}. 
For each requirement unit, the annotators first examine whether a code artifact contains explicit implementation evidence that is substantively related to the requirement. 
If such evidence is identified, the annotators further assess whether the artifact directly realizes the mechanism, constraint, condition, or assurance expressed in the requirement.

A requirement--code pair is labeled as \textit{Positive} only when direct realization is present. 
A pair is labeled as \textit{Negative} when the artifact is topically or operationally related to the requirement but does not directly realize it, or provides only partial or insufficient evidence. 
A requirement unit is labeled as \textit{Missing} when no explicit implementation evidence can be identified in the current codebase under file-level granularity.

\subsection{Example Annotations}
Table~\ref{tab:examples} presents four representative requirement--code pairs to illustrate the annotation protocol. The examples cover four typical outcomes: a positive NFR link with explicit implementation evidence, a positive FR link grounded in functional logic, a negative case where a candidate code unit is related but does not directly implement the requirement, and a missing case where no explicit implementation evidence is found in the current codebase. Together, these examples demonstrate the application of the \textit{direct implementation responsibility} principle.
\begin{table}[t]
\centering
\caption{Representative Annotated Requirement--Code Pairs}
\label{tab:examples}
\scriptsize
\setlength{\tabcolsep}{2.2pt}
\renewcommand{\arraystretch}{1.12}
\begin{tabular}{
@{}
>{\raggedright\arraybackslash}p{1.35cm}
>{\raggedright\arraybackslash}p{2.65cm}
>{\centering\arraybackslash}p{0.65cm}
>{\raggedright\arraybackslash}p{2.15cm}
>{\raggedright\arraybackslash}p{1.20cm}
>{\raggedright\arraybackslash}p{2.55cm}
@{}}
\toprule
\textbf{Req. ID} & \textbf{Requirement} & \textbf{Type} & \textbf{Code Unit} & \textbf{Label} & \textbf{Rationale} \\
\midrule
NFR-Sec-07
& MQTT payload shall be encrypted using SM4.
& NFR
& \texttt{MqttService\allowbreak.java}
& Positive
& SM4-based payload decryption implemented. \\

FR-Func-04
& Provide interface for on-chain evidence storage.
& FR
& \texttt{BlockChainCtrl\allowbreak.java}
& Positive
& API for blockchain evidence storage. \\

NFR-Sec-05
& Communication shall prevent man-in-the-middle attacks.
& NFR
& \texttt{MqttConfig\allowbreak.java}
& Negative
& No TLS/SSL; security requirement not satisfied. \\

NFR-Perf-01
& Simple queries shall respond within 1 second.
& NFR
& \texttt{None}
& Missing
& No explicit performance enforcement logic. \\
\bottomrule
\end{tabular}
\end{table}

\section{Dataset Observations}
\label{sec:dataset_observations}
\subsection{Traceability Challenges with NFRs}
\begin{table}[t]
\centering
\caption{Overview of the Curated Traceability Subset}
\label{tab:overview}
\footnotesize
\setlength{\tabcolsep}{5pt}
\renewcommand{\arraystretch}{1.08}
\begin{tabular}{
@{}
>{\raggedright\arraybackslash}p{4.0cm}
>{\centering\arraybackslash}p{1.0cm}
>{\raggedright\arraybackslash}p{4.2cm}
@{}}
\toprule
\textbf{Item} & \textbf{Count} & \textbf{Percentage / Notes} \\
\midrule
\multicolumn{3}{@{}l}{\textbf{Dataset size}} \\
Requirement units & 38 & 100\% \\
FR units & 11 & 28.9\% of requirements \\
NFR units & 27 & 71.1\% of requirements \\
Code units & 22 & File-level artifacts \\

\midrule
\multicolumn{3}{@{}l}{\textbf{Traceability outcomes}} \\
Positive trace instances & 18 & 47.4\% of requirements \\
Negative trace instances & 11 & 28.9\% of requirements \\
Missing trace instances & 9 & 23.7\% of requirements \\
FR positive instances & 9 & 81.8\% of FRs \\
NFR positive instances & 9 & 33.3\% of NFRs \\

\midrule
\multicolumn{3}{@{}l}{\textbf{NFR category distribution}} \\
Security NFR units & 9 & 33.3\% of NFRs \\
Performance NFR units & 8 & 29.6\% of NFRs \\
Other NFR units & 10 & 37.0\% of NFRs \\

\midrule
\multicolumn{3}{@{}l}{\textbf{Security NFR outcomes}} \\
Security positive & 3 & 33.3\% of Security NFRs \\
Security negative & 6 & 66.7\% of Security NFRs \\
Security missing & 0 & 0.0\% of Security NFRs \\
\bottomrule
\end{tabular}

\vspace{0.8mm}
\begin{minipage}{0.92\textwidth}
\footnotesize
\textit{Note}: Other NFR units include Reliability, Data Integrity, Auditability, Scalability, and Maintainability.
\end{minipage}
\end{table}

Table~\ref{tab:overview} summarizes the curated subset used in this study. Non-functional requirements constitute a substantial portion of the extracted requirement units: 27 out of 38 units (71.1\%) are labeled as NFRs, whereas only 11 units (28.9\%) are labeled as FRs. This distribution suggests that, at least in this curated subset, the challenge of NFR traceability does not arise from a lack of NFR-related content in the source documents.

A clearer contrast emerges when the traceability outcomes are examined. Among the 11 FR units, 9 are labeled as \textit{Positive}, corresponding to 81.8\% of the FR subset. In contrast, among the 27 NFR units, only 9 are labeled as \textit{Positive}, yielding a positive rate of 33.3\%. The remaining 18 NFR units are evenly divided between \textit{Negative} and \textit{Missing} cases. This gap is consistent with the view that the main difficulty of NFR tracing lies not in requirement presence, but in identifying explicit and defensible implementation evidence under the criterion of direct implementation responsibility.

One possible explanation is the mismatch in abstraction between many NFR statements and their implementation evidence. In the current project, a considerable portion of NFRs are expressed as system-level quality goals or operational constraints, whereas the relevant implementation evidence, when present, is often partial, implicit, or distributed across multiple artifacts. Some requirements remain at the level of architectural intent without being realized through a single explicit file-level artifact, while others depend on framework behavior or runtime configuration rather than directly visible business logic. As a result, many candidate artifacts are related to the requirement but do not satisfy the criterion for a positive trace link at the current granularity. This observation is broadly consistent with prior work on the difficulty of tracing NFRs due to their implicit and distributed implementation characteristics~\cite{cleland2005toward,mahmoud2016detecting}.

\subsection{Heterogeneity of NFR Links}
\begin{figure}[t]
\centering
\begin{tikzpicture}
\begin{axis}[
    ybar,
    bar width=9pt,
    width=0.92\textwidth,
    height=0.48\textwidth,
    ymin=0,
    ymax=5,
    ylabel={Positive Link Count},
    symbolic x coords={Controller,Service,Config,Utility,Contract},
    xtick=data,
    xticklabels={Controller,Service,{Config\\Component},Utility,Contract},
    xticklabel style={
        font=\scriptsize,
        align=center
    },
    ytick={0,1,2,3,4,5},
    ymajorgrids=true,
    grid style={dashed, gray!25},
    enlarge x limits=0.12,
    axis x line*=bottom,
    axis y line*=left,
    tick style={black},
    line width=0.6pt,
    legend style={
        font=\scriptsize,
        draw=none,
        fill=none,
        at={(0.98,0.98)},
        anchor=north east,
        legend columns=2,
        /tikz/every even column/.append style={column sep=4pt}
    },
    nodes near coords,
    every node near coord/.append style={
        font=\scriptsize
    },
    visualization depends on={y \as \rawy},
    nodes near coords={
        \pgfmathtruncatemacro{\checkzero}{\rawy==0}
        \ifnum\checkzero=1
        \else
        \pgfmathprintnumber{\rawy}
        \fi
    }
]
\addplot[
    draw=blue!70!black,
    fill=blue!20
] coordinates {
    (Controller,2)
    (Service,4)
    (Config,2)
    (Utility,0)
    (Contract,1)
};

\addplot[
    draw=red!70!black,
    fill=red!20
] coordinates {
    (Controller,2)
    (Service,3)
    (Config,1)
    (Utility,1)
    (Contract,2)
};

\legend{FR Positive,NFR Positive}
\end{axis}
\end{tikzpicture}
\caption{Distribution of positive trace links by implementation role.}
\label{fig:impl_role}
\end{figure}
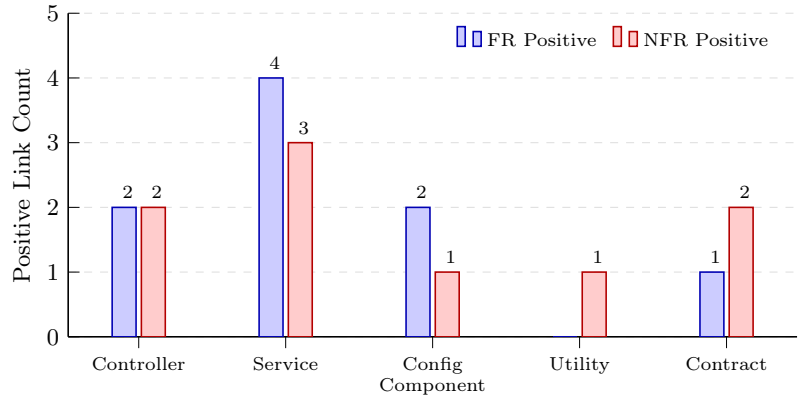

Beyond the lower rate of positive links, NFR traceability in the current subset also exhibits greater implementation heterogeneity than FR traceability. 
Fig.~\ref{fig:impl_role} shows the distribution of positive trace links by implementation role, where the role labels are derived from file-level artifact metadata. A clear pattern is that FR positive links are concentrated in relatively explicit business-oriented locations, especially controller and service artifacts. Together, these two roles account for 6 out of 9 FR positive instances.

In contrast, NFR positive links are distributed across a broader range of implementation roles. 
Although some NFR evidence also appears in controllers and services, additional positive links are found in utility code, configuration artifacts, components, and smart contracts.
This pattern is consistent with the observation that FRs are often realized through relatively direct execution paths, whereas NFRs are more likely to be enforced through supporting mechanisms, runtime configuration, validation logic, or contract-level constraints. 
In particular, utility-level positive evidence appears only for NFRs in the current subset, and contract artifacts contribute proportionally more positive evidence for NFRs than for FRs. This suggests that some NFRs are implemented through low-level supporting mechanisms or on-chain constraints rather than through explicit business functions.

This heterogeneity has practical implications for traceability analysis. First, relevant NFR evidence is less likely to be concentrated in a single obvious artifact, making it harder to locate direct implementation support. Second, the implementation roles associated with NFRs are more diverse and often less lexically aligned with requirement text than those associated with FRs. As a result, retrieval methods that perform reasonably well in FR-oriented settings may be less effective for NFR tracing, especially when relevant evidence resides in configuration, utility, or contract artifacts that are only weakly connected to the surface wording of the requirement.

\subsection{Challenges with Security NFRs}
As summarized in Table~\ref{tab:overview}, security-related NFRs form the largest NFR category in the current subset, accounting for 9 out of 27 NFR units (33.3\%). At the same time, they exhibit a relatively low rate of positive trace links: only 3 of the 9 security NFR units are labeled as \textit{Positive}, while the remaining 6 are labeled as \textit{Negative}. Compared with other requirement types, security-related NFRs are more often specified as mandatory system assurances, yet less likely to correspond to a single explicit and self-contained implementation artifact.

This difficulty can be attributed to several factors at the requirement level. First, many security-related NFRs are expressed in broad and high-level terms, such as secure transmission, access control, or protection against common attacks. While semantically clear, these descriptions do not always correspond to a single operational mechanism that can be directly matched to a file-level code unit. Second, even when related artifacts exist, they often implement only local mechanisms rather than the full assurance expressed in the requirement. As a result, requirement--code pairs may appear topically related while still failing to satisfy the criterion for a positive trace link.

At the implementation level, security evidence often does not appear as a single, clearly identifiable unit as in typical FR implementations. In the current project, relevant evidence may reside in configuration classes, communication settings, or utility-level processing logic rather than in clearly identifiable business-layer functionality. This leads to two additional challenges: (1) the observable evidence is often partial and does not fully justify the requirement, and (2) some expected protections rely on configuration choices or framework-level behavior rather than explicit code.

\subsection{Case Studies and Insights}
To ground the above observations, we examine two representative cases from the annotated subset, illustrating positive and negative traceability outcomes for security-related NFRs.

\begin{figure}[!t]
\centering

\begin{minipage}{0.76\textwidth}
\centering
\includegraphics[width=\linewidth]{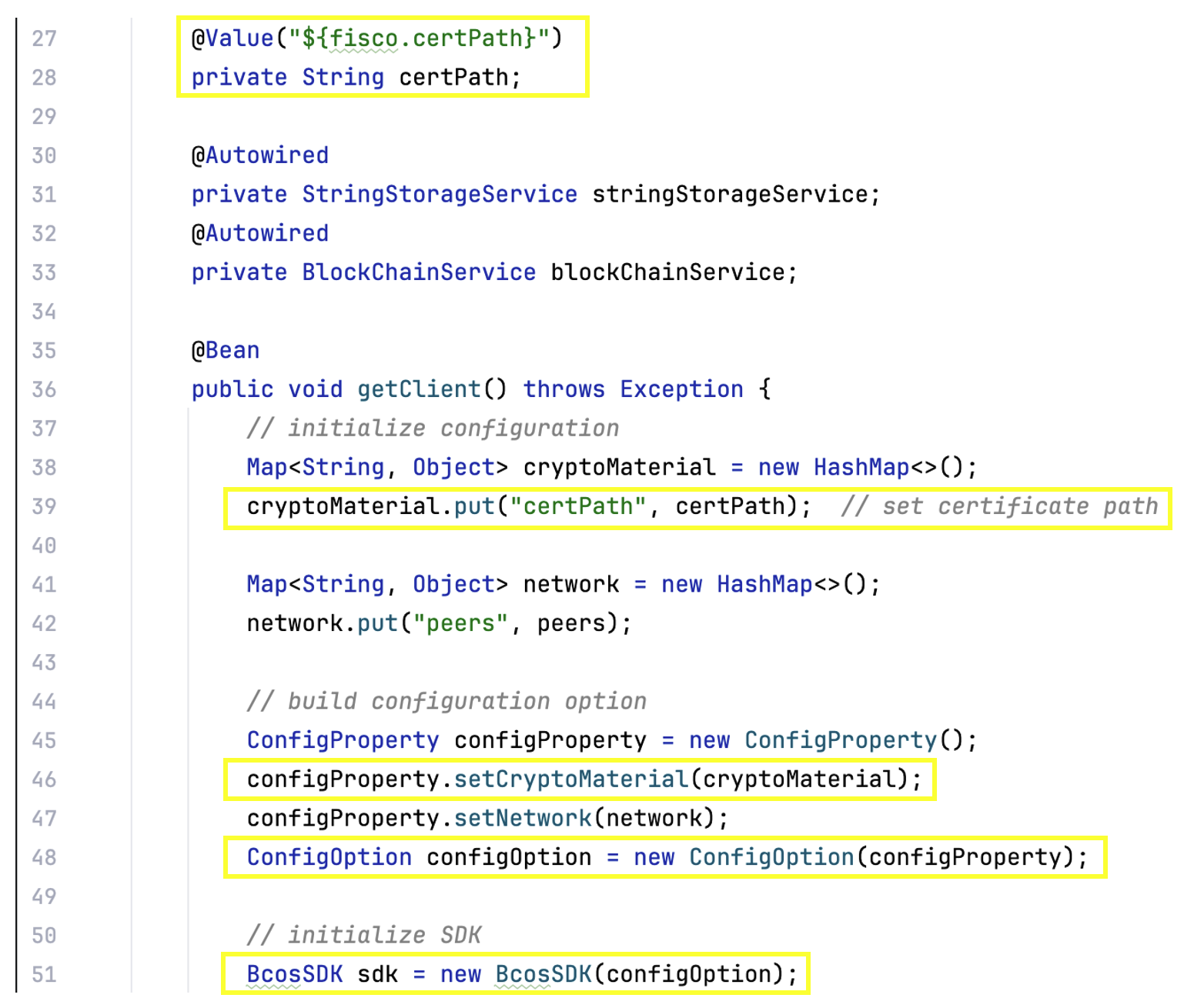}
\\[-0.5em]
\footnotesize
(a) Positive evidence for \textit{NFR-Sec-09} in \texttt{FiscoBcosConfig.java}.
\end{minipage}

\vspace{0.25em}

\begin{minipage}{0.76\textwidth}
\centering
\includegraphics[width=\linewidth]{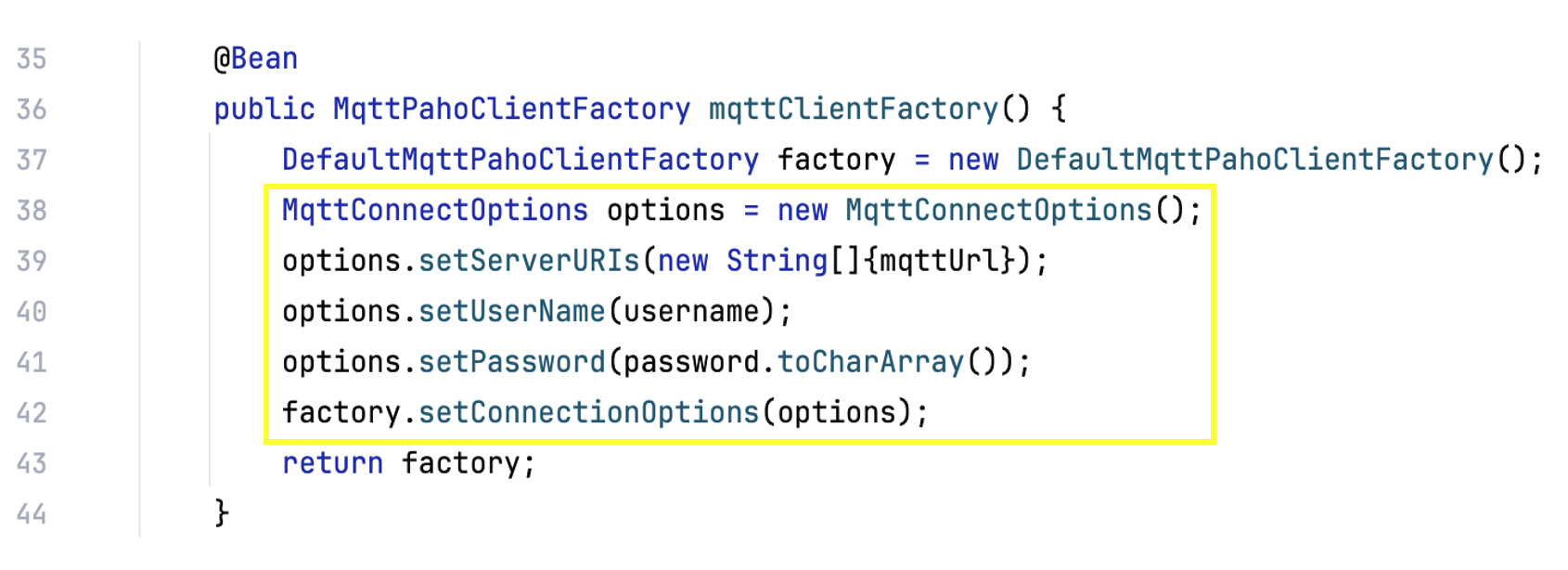}
\\[-0.5em]
\footnotesize
(b) No explicit TLS/SSL enforcement in \texttt{MqttConfig.java}.
\end{minipage}

\vspace{-0.3em}
\caption{Representative security NFR traceability cases.}
\label{fig:security_cases}
\end{figure}

\textbf{Case A: Traceable security NFR with explicit implementation evidence.}
A representative positive case is \textit{NFR-Sec-09}, which requires that IoT devices be authenticated through digital certificates or identity credentials during access. As shown in Fig.~\ref{fig:security_cases}(a), \texttt{component/FiscoBcosConfig.java} explicitly injects the certificate path through \texttt{certPath}, incorporates it into the cryptographic material via \texttt{cryptoMaterial.put("certPath", certPath)}, and propagates this configuration into SDK initialization through \texttt{setCryptoMaterial\allowbreak(...)} and \texttt{new BcosSDK\allowbreak(configOption)}. This artifact provides explicit implementation evidence of a certificate-based credential mechanism. Under the criterion of direct implementation responsibility, it can therefore be labeled as a positive trace target.

\textbf{Case B: Related communication artifact without explicit secure-transport enforcement.}
A contrasting case is \textit{NFR-Sec-05}, which requires that system communication adopt encryption mechanisms such as TLS/SSL to prevent man-in-the-middle attacks. As shown in Fig.~\ref{fig:security_cases}(b), 
\texttt{MqttConfig.java} is responsible for MQTT communication setup: it constructs \texttt{MqttConnectOptions},
assigns the broker URI through \texttt{setServerURIs(...)}, and configures connection parameters based on username and password.
However, the artifact does not explicitly initialize TLS/SSL-related mechanisms such as SSL context or certificate-based transport configuration. Under the criterion of direct implementation responsibility, it is therefore relevant to the requirement but cannot be labeled as a positive trace target.

\textbf{Insight.}
These cases show that positive trace links rely on explicit and localized implementation evidence, while artifacts that are only operationally related may still lack sufficient support for a security requirement.

\section{Experiments}
\subsection{Task Definition}
To further examine the observations from Section~\ref{sec:dataset_observations}, we formulate Requirements-to-Code trace recovery as a retrieval task in the constructed empirical setting.
Given a requirement unit as a query and a set of candidate code units, each retrieval method assigns a relevance score to every candidate artifact and ranks all candidates accordingly. The ranked list is then used to recover trace links for evaluation.

For the industrial curated subset, each requirement query is ranked against the full set of 22 file-level code units. A code unit is considered \textit{relevant} only when the corresponding requirement--code pair is labeled as \textit{Positive} in the adjudicated trace set. Requirement units with no adjudicated positive code artifact are excluded from ranking-based retrieval evaluation, and are analyzed separately in the descriptive analysis and discussion.

In the current study, the task is examined under three settings on the industrial curated subset: \textit{FR-to-Code}, \textit{NFR-to-Code}, and \textit{Security NFR-to-Code}. The first two settings compare the overall retrieval difficulty of functional and non-functional requirements, while the third focuses on security-related NFRs as a representative category for more fine-grained analysis.

\subsection{Settings}
\subsubsection{Datasets}
We conduct a baseline evaluation of the task on two datasets. The primary dataset is the industrial curated subset constructed in this study, which contains 38 requirement units, 22 file-level code units, and manually adjudicated trace links covering both FR and NFR instances. Following the evaluation protocol above, only requirement units with at least one positive code artifact are included in retrieval evaluation. Based on the current annotated subset, this yields 9 FR queries, 9 NFR queries, and 3 Security NFR queries.

As an auxiliary public reference, we additionally include iTrust, a widely used requirements-to-code traceability benchmark containing 131 requirement artifacts and 226 Java code artifacts with established trace links. Since iTrust mainly represents a conventional FR-oriented setting, it is used only as an external reference point rather than the primary target of analysis or a directly controlled benchmark comparison.

\subsubsection{Methods}
We evaluate four classical retrieval-based baselines: TF-IDF~\cite{salton1988term}, BM25~\cite{robertson2009probabilistic}, LSI~\cite{deerwester1990indexing}, and WMD~\cite{kusner2015word}.
TF-IDF and BM25 represent lexical matching methods, while LSI and WMD capture broader semantic similarity. For the industrial curated subset, each requirement query is matched against all 22 file-level code units, and each method produces a ranked list of candidate artifacts according to the computed relevance scores. The same retrieval-based setting is applied to iTrust.

\subsubsection{Metrics}
We report Precision, Recall, F1, and Mean Reciprocal Rank (MRR) to evaluate retrieval effectiveness. For each requirement query, the ranked results are compared against the set of adjudicated relevant code artifacts. Precision, Recall, and F1 are computed at the query level and then macro-averaged across all evaluated queries. MRR is computed based on the rank position of the first relevant code artifact for each query and is also macro-averaged across queries.

\subsection{Results}
Table~\ref{tab:combined_results} summarizes the retrieval results across the industrial curated subset and the iTrust reference dataset. On the industrial subset, FR-to-Code consistently outperforms NFR-to-Code across all four baselines and all reported metrics. BM25 achieves the strongest overall performance, while LSI performs worst overall.

Security NFR queries yield lower scores than the overall NFR setting across all baselines, making them the most difficult subset in the current experiment. On this subset, BM25 again performs best, while WMD is slightly stronger than TF-IDF on several metrics.

On iTrust, all methods achieve higher and more stable results than on the industrial NFR and Security NFR settings. BM25 again ranks first overall.
\begin{table}[t]
\centering
\caption{Retrieval Results Across Industrial and Reference Settings}
\label{tab:combined_results}
\footnotesize
\setlength{\tabcolsep}{5pt}
\renewcommand{\arraystretch}{1.08}
\begin{tabular}{llcccc}
\toprule
\textbf{Setting} & \textbf{Metric} & \textbf{TF-IDF} & \textbf{BM25} & \textbf{LSI} & \textbf{WMD} \\
\midrule
\multirow{4}{*}{FR-to-Code}
& Prec. & 0.71 & 0.74 & 0.61 & 0.68 \\
& Rec.  & 0.62 & 0.68 & 0.57 & 0.63 \\
& F1    & 0.66 & 0.71 & 0.59 & 0.65 \\
& MRR   & 0.72 & 0.79 & 0.66 & 0.74 \\
\midrule
\multirow{4}{*}{NFR-to-Code}
& Prec. & 0.43 & 0.47 & 0.35 & 0.41 \\
& Rec.  & 0.34 & 0.40 & 0.30 & 0.38 \\
& F1    & 0.38 & 0.43 & 0.32 & 0.39 \\
& MRR   & 0.45 & 0.51 & 0.40 & 0.49 \\
\midrule
\multirow{4}{*}{Security NFR}
& Prec. & 0.33 & 0.39 & 0.28 & 0.35 \\
& Rec.  & 0.27 & 0.32 & 0.24 & 0.31 \\
& F1    & 0.30 & 0.35 & 0.26 & 0.33 \\
& MRR   & 0.39 & 0.46 & 0.35 & 0.44 \\
\midrule
\multirow{4}{*}{iTrust}
& Prec. & 0.61 & 0.62 & 0.57 & 0.59 \\
& Rec.  & 0.60 & 0.61 & 0.56 & 0.59 \\
& F1    & 0.60 & 0.62 & 0.57 & 0.59 \\
& MRR   & 0.70 & 0.72 & 0.67 & 0.70 \\
\bottomrule
\end{tabular}
\end{table}

\subsection{Discussion}
The results are consistent with the observations from Section~\ref{sec:dataset_observations}.
First, FR-to-Code traceability is consistently easier than NFR-to-Code tracing. 
This pattern reflects a structural difference in how requirements are implemented: FRs are more often realized through explicit business-oriented artifacts, whereas NFR evidence is more likely to be distributed across heterogeneous artifacts and is therefore often weakly localized in code. 
As a result, NFR trace recovery is hindered not only by weaker lexical alignment, but also by the lack of clearly identifiable implementation anchors. 
This structural property is also reflected in the baseline results: lexical methods such as BM25 and TF-IDF remain competitive, while the classical semantic baselines evaluated here do not substantially alleviate the difficulty of NFR trace recovery.

Security-related NFRs are particularly difficult in this regard. Compared with the broader NFR subset, they yield consistently weaker retrieval results across baselines. This suggests that security-related NFRs are especially likely to be expressed as high-level assurances, while their implementation is indirect, configuration-dependent, or distributed across multiple supporting locations. The difficulty therefore lies not only in terminology mismatch, but also in the limited visibility of explicit code-level evidence.

An additional observation concerns requirement units labeled as \textit{Missing}, i.e., cases for which no explicit implementation evidence could be identified in the current codebase. These requirements are excluded from ranking-based evaluation because they contain no relevant target artifact. Nevertheless, the observed score patterns suggest that such requirements often receive uniformly low retrieval scores across candidate artifacts. This indicates that retrieval confidence may provide a weak signal for the absence of explicit implementation evidence. Beyond recovering existing trace links, low-ranked and low-confidence results may also help identify requirement units whose implementation support is incomplete, indirect, or not explicitly localized at the current file-level granularity.

\section{Related Work}
Requirements traceability has been extensively studied for recovering links between requirements and downstream software artifacts, such as source code, design models, and test cases ~\cite{cleland2014software}. 
Classical approaches in this area are largely based on information retrieval, similarity ranking, and artifact matching~\cite{hayes2003improving,borg2014recovering}, 
and public datasets such as iTrust have been widely used to evaluate trace recovery methods~\cite{hey2021improving,mader2012assessing}. 
However, these benchmarks and evaluations are predominantly FR-oriented, and therefore provide limited evidence on how trace recovery behaves for non-functional requirements in heterogeneous industrial settings.

Research on NFRs has more often focused on requirement-side tasks, including NFR elicitation, detection, extraction, identification, and classification~\cite{abad2017works}. 
Prior studies have investigated how to recognize quality attributes such as security, performance, and maintainability from requirement text using rule-based and learning-based approaches ~\cite{rahman2023non}. 
While these studies focus on NFR understanding at the requirement-text level, they do not address how NFRs are grounded in implementation artifacts. Explicit NFR-to-Code traceability remains underexplored~\cite{mahmoud2016detecting}, especially in settings where implementation is distributed across heterogeneous artifacts rather than localized.

Recent work has begun to explore the use of large language models (LLMs) in traceability, including retrieval enhancement and cross-artifact matching~\cite{ge2025cross,wang2026r2code,mao2025hybrid}. Related studies also apply LLMs to user-story elaboration and design-model derivation~\cite{li2024simac,li2024llm}, indicating their potential to bridge semantic gaps between artifacts~\cite{guo2025natural}.
Nevertheless, evidence mainly comes from conventional FR-oriented benchmark settings. Whether such techniques can effectively support industrial NFR-to-Code traceability remains unclear, especially when the difficulty lies not only in semantic mismatch but also in distributed and weakly localized implementation evidence.
Therefore, this work focuses on an industrial NFR-to-Code setting and provides an empirical comparison between FR and NFR traceability under heterogeneous implementation conditions.

\section{Threats to Validity}
\textbf{Internal validity.}
This study relies on manual requirement decomposition, requirement typing, and trace link annotation, and some degree of subjectivity therefore remains unavoidable. To mitigate this threat, we designed and applied a structured annotation protocol. 
Two annotators independently labeled the requirement--code pairs following the same criteria, and disagreements were resolved through discussion and adjudication; future extensions will further quantify annotation agreement to strengthen the reliability analysis.
Nevertheless, borderline cases may still affect how NFR categories and \textit{Positive}/\allowbreak\textit{Negative}/\allowbreak\textit{Missing} labels are assigned.
In addition, the study adopts file-level code units, which improve consistency and feasibility but may under-represent requirements whose implementation evidence is distributed across multiple files or finer-grained program elements.

\textbf{External validity.}
The current dataset is constructed from a single industrial blockchain-IoT project and a manually curated subset with a relatively small sample size. As a result, the observed traceability patterns may still be influenced by project-specific factors such as domain characteristics, coding conventions, and documentation style. To partially mitigate this limitation, we additionally include iTrust as an external FR-oriented reference dataset. However, publicly available datasets that explicitly support NFR-to-Code traceability remain scarce, especially for heterogeneous industrial settings.

\section{Conclusion}
This paper presents an empirical study of NFR-to-Code traceability based on a real-world blockchain-IoT project. We design an annotation protocol for constructing trace links across heterogeneous artifacts, build a manually curated FR/NFR-to-Code trace subset, and evaluate representative retrieval-based baselines on this dataset.

The study shows that NFR traceability is fundamentally more challenging than FR traceability, with security-related NFRs being particularly difficult. More importantly, the difficulty does not stem from the absence of requirements, but from the fact that their implementation evidence is distributed across heterogeneous artifacts and not clearly localized in code. This observation has direct implications for quality-sensitive systems, where requirements related to security, reliability, and performance play a central role. It suggests that improving traceability in such settings requires methods that go beyond conventional lexical matching and explicitly account for distributed and implicit implementation evidence. Future work will extend the dataset, report quantitative inter-annotator agreement, and evaluate modern embedding-based, neural, and LLM-based traceability techniques tailored to distributed and weakly localized NFR evidence.
\section*{Acknowledgment}
The work described in this paper was substantially supported by Hong Kong Metropolitan University Research Grant (No. RD/2025/1.21) and the Faculty Development Scheme of the University Grants (No. UGC/FDS16/E25/25).
This work is also supported in part by the General Research Fund of the Research Grants Council of Hong Kong and the research funds of the City University of Hong Kong (6000871, 9229109, 9229029, 9229192).

\medskip
\noindent\textbf{Disclosure of Interests.}
The authors have no competing interests to declare that are relevant to the content of this article.

\bibliographystyle{splncs04}
\bibliography{reference}
\end{document}